\documentclass[12pt]{article}
\usepackage[dvips]{color}
\usepackage{epsfig}
\usepackage{graphicx}
\usepackage{amsmath}

\begin{document}
\begin{center}
\Large{\bf The relation between non-commutative and Finsler geometry in Horava-Lifshitz black holes }\\
\small \vspace{1cm} {\bf Z. Nekouee
\footnote{z.nekouee@stu.umz.ac.ir}},\quad{\bf J. Sadeghi
\footnote{pouriya@ipm.ir}}\quad and\quad{\bf M. Shokri \footnote{mehdi.shokri@uniroma1.it}}\\
\vspace{0.5cm}$^{1}${\it Department of Mathematic, Faculty of Mathematic,\\
University of Mazandaran, P. O. Box 47416-95447, Babolsar, Iran}\\
\vspace{0.5cm}$^{2}${\it Department of Physics, Faculty of Basic Sciences,\\
University of Mazandaran, P. O. Box 47416-95447, Babolsar, Iran}\\
\vspace{0.5cm}$^{3}${\it Physics Department and INFN,\\
Universit`a di Roma “La Sapienza”, Ple. Aldo Moro 2, 00185, Rome, Italy}\\
\end{center}\vspace{1.2cm}
\date{}

\begin{abstract}
In this paper we employ the Horava-Lifshitz black holes solutions and obtain the corresponding
Hamiltonian. It helps us to take new variables and it will be written by
harmonic oscillator form. This leads us to apply non-commutative geometry
to the new Hamiltonian and obtain the corresponding Lagrangian. And then, we take some
information from Finsler geometry and write the Lagrangian of the different kinds of Horava-Lifshitz black holes. We
show that the corresponding Lagrangian in non-commutative and Finsler geometry for above mentioned black
holes completely coincidence together with some specification of parameters. But in case of rotation, the place of center of mass energy completely different, so the particle goes to inside of black hole rapidly without falling into singularity. So in that case, two Lagrangians cover each other at $0<r<r_h.$ 
\end{abstract}
\newpage
\section{Introduction}
General relativity ($GR$) as a classical theory of gravity cannot explain very early universe. Hence, it is necessary to generate a new gravitational theory in the context of quantum physics which is known quantum gravity ($QG$) [1,2]. This theory is one of the greatest mysteries of the universe because it suffers from many fundamental ambiguities. Therefore, in order to explain it, there are many versions of this theory in cosmological papers. String theory in addition to solve the unification problem, attempts to illustrate $QG$  by means of holography principle [3-5]. On the other hand, some people believe that we can fabricate $QG$ in the context of $GR$ by considering some quantum corrections but there are many versions in the literature [6]. One of the most significant models proposed by Horava which is known Horava-Lifshtiz ($HL$) gravity [7,8]. Such theory does not have the full diffeomorphism invariance of $GR$, so a different scaling symmetry is allowed in the $UV$ and the theory accepts renormalizable couplings with higher order spatial derivatives. According to corresponding theory, we are faced with some anisotropic scaling relation $x \rightarrow bx$ and $t \rightarrow b^{z}t$ where $b$ is a constant and $z$ is responsible to measure the degree of anisotropy of space and time coordinates. $HL$ gravity is based on the basic assumption that the Lorentz symmetry is fundamentally broken at high scales of energy and restores only in the infrared ($IR$) limit. In contrast, some people believe that for each finite energy scale the Lorentz symmetry is broken and the $HL$ gravity admits Lorentz symmetry preserving preferred global time foliation of the space-time. This symmetry can be related to the standard Lorentz transformations by the frame dependent change of synchronization to the Einstein one [9]. Another interesting feature of this theory is that $HL$ gravity can supply the minimal holographic dual for field theories which are Lifshitz-type with anisotropic scaling and dynamical exponent z [10]. Also, one can find the hydrodynamics of $HL$ gravity in [11]. $HL$ gravity has applied in a wide range of cosmological scenarios from small to large scale. Also, it has engaged to explain the physics of black holes in many literatures. Some solutions of charged $HL$ black holes were reported in [12]. Also, the generalized form of $BTZ$ black holes in the context of $HL$ gravity can be found in [13]. In Ref. [14], the hydrodynamics of $HL$ black holes was interpreted. Moreover, people studied electromeganetic static spacetimes in the non-relativisitc general covariant theory of $HL$ gravity [15]. A detailed investigation of $HL$ black holes were presented in [16]. Recently, an interesting review has carried out by Anzhong Wang about $HL$ gravity and its developments in different areas [17]. Although, the investigation of $HL$ black holes seems be simple asymptotically, but it can be very difficult when we have interested to present a analytic solution [18]. As an example, a $4D$ topological black hole with the dynamical exponent $z = 2$ which is Lifshitz asymptotically [19]. An analytic black hole solution with $z = 2$ that asymptotes a planer Lifshitz background was found in $4D$ space-times [20], and its numerical solutions were explored in [21,22]. The analytic examples of Lifshitz black holes in higher dimensions were investigated in [23]. From new massive gravity in $3D$ space-times [24], the Lifshitz black hole was derived with $z = 3$ [25]. About the thermodynamics of $HL$ black holes, the Ref. [26] was dedicated to calculate thermodynamics quantities in this kinds of black holes. Moreover, people have done some attempts on the modified type of $HL$ black holes [27].\\
Another important issue here is associated to non-commutative ($NC$) geometry which can be realized in context of quantum physics [28]. When we want to move from classical to quantum mechanics, the space also will be transited from classic to quantum space where there are some commutative relations between operators in the Hilbert space. Although this approach of space has many applications in physics. We note that here, there are some areas of physics need to change the definition of space. Hence, we can focus our attention to $NC$ geometry as a special kind of geometry which has many applications in physic. One of the most important applications of this geometry can be found to generate a $QG$ theory. In holographic models of $QG$, we deal with a kind of non-commutativity between string and field theories [29]. In loop quantum gravity ($LQG$), the space is NC and discrete, in contrast the canonical quantum gravity ($CQG$) model has commutative and continues space. Another application of $NC$ geometry is associated to a class of $CQG$ models with $NC$ space which can be imposed to our scenarios by deformed space-time. Our recent attention to this kind of deformation can be followed in the previous works for dark energy, massive gravity and physics of black holes [30-32].\\
On the other hand, since some part of the present work is based on Finsler geometry, we give some explanation about this geometry. Finsler geometry was first applied in gravitational theory, and this application lead to corrections to observational results predicted by general relativity [33-41]. Although Riemannian geometry as a mathematical foundation of $GR$ can obtain remarkable achievements in order to explain gravitation in large scale structure.  But sometimes we need to the generalized form of the Riemannian geometry to illustrate our gravitational situations. Therefore, we focus our attention to Finsler geometry. From mathematical approach length of a vector $\|x\|$ in a manifold $M$ endowed by a metric $g_{ij}(x)$ in Riemannian geometry is given by the quadratic form,
\begin{equation}
\|x^{2}\|=g_{ij}(x)x^{i}x^{j}
\end{equation}
where $i$ and $j$ denote to spatial components. Moreover, Finsler geometry can present a more general method to determine
the norms of the vectors. Because of this reason, Finsler geometry is a generalization of Riemaniann geometry. In this geometry, the length of the vectors are determined by a general method which is not restricted by Riemaniann definition of length in terms of square root of the quadratic form. According to the mathematical definition of this geometry, we can introduce a $N$ dimensional manifold $M$ which is supplied by a positive scalar function $F(x,y)$
\begin{equation}
F(x,y):TM\rightarrow[0,\infty)
\end{equation}
where $x=x^{i}=(x^{0}, ...., x^{n})\in M$, $y=y^{i}=(y^{0}, ...., y^{n})\in T_{x}M$ and $TM$ is the tangent bundle.\\
Finsler metric function have to satisfy three properties: The first is regularity, it means that $F(x,y)$ is differentiable on the slit tangent bundle $TM\backslash0$. The second is positive homogeneity, e.g. $F(x,y)$ is homogenous function of degree one in $y$ as $F(x,\lambda y)=\lambda F(x,y)$ for any number of positive $\lambda$. The third is Strong convexity, it can be realized that the quadratic form $\frac{\partial^{2}F(x,y)}{\partial y^{i}\partial y^{j}}y^{i}y^{j}$ is assumed to be positive definite for all variables [42-44]. In this paper, we study some important physical Lagrangians and attempt to connect Finsler geometry and $NC$ geometry (deformed space-time) together. As we know, all Lagrangians are not able to satisfy the regularity condition of Finsler geometry. On the other hand, the majority of Lagrangians in gauge theory cannot supply the convexity condition of Finsler geometry. Hence, we restrict our analysis to a class of gauge theory Lagrangian which are equipped to weak regularity and positive homogeneity conditions of Finsler geometry.\\
Any physical and mathematical Lagrangian systems of finite degree of freedom can be reformulated
in Finsler manifolds without changing their physical contents [45,46]. In that case,
the action functional is given by the integral of the Finsler metric which is made from the
Lagrangian. Then the variational principle becomes geometric and independent of parametrisation,
which we will call covariant. From the point of view of a physicist, especially when
thinking about the Lagrangian formulation, we are inclined to define a non-linear connection
not on a line element space $TM^{0}$, but directly on the point manifold $M$ [47]. So, in this paper
we make the Lagrangian from $HL$ black holes solutions and obtain the corresponding
Hamiltonian. In such approach, we take some new variables to write the Hamiltonian
in form of harmonic oscillator Hamiltonian. This give us motivation to apply $NC$ geometry to the new Hamiltonian and obtain the corresponding Lagrangian. And then, we employ some information from Finsler metric and Lagrangian for the different kinds of $HL$ black holes.
We show that the corresponding Lagrangian in $NC$ geometry for the mentioned black holes completely coincidence with Lagrangian from Finsler geometry with some specification of parameters.
\section{The analysis}
In this section, firstly we consider four types of $HL$ black holes in the context of $HL$ gravity. Then, we apply $NC$ geometry in order to modify the phase space for each kind of $HL$
black holes, separately. In the next step, the set up will be investigated in the Finsler geometry regime.
\subsection{The charged Horava-Lifshitz black hole}
According to Ref. [48], the form of action for the gravitational section in $HL$ gravity can be found as,
\begin{equation}
S_{g}=\frac{1}{2\kappa^{2}}\int dtdrd^{d}x\sqrt{-G}(K_{ab}K^{ab}-\lambda K^{2}+\beta(R-2\Lambda)+\frac{\alpha^{2}}{2}\frac{\nabla_{a}N\nabla^{a}N}{N^{2}}),
\end{equation}
where for $\lambda=\beta=1$ and $\alpha=0$, the form of action is reduced to $GR$. In the analysis, we can engage a general type of metric as,
\begin{equation}
ds^{2}=-N^{2}dt^{2}+g_{ij}(dx^{i}-N^{i}dt)(dx^{j}-N^{j}dt),
\end{equation}
where $N$, $N^{i}$ and $g_{ij}$ are the lapse function, the shift vector and the metric of the space-like
hypersurface, respectively. In that case, we can have formula $\sqrt{-G} = \sqrt{g}N$,
 $K_{ab}=\frac{1}{2N}(\partial_{t}g_{ab}-\nabla_{a}N_{b}-\nabla_{b}N_{a})$ and $K=g^{ab}K_{ab}$. In order to investigate $HL$ black holes, we consider Lorentz-violating electromagnetism field as matter sector [49],
\begin{eqnarray}
S_{m}=-\frac{1}{2\kappa^{2}}\int dtdrd^{d}x\sqrt{g}N(\frac{2}{N^{2}}g^{ij}(F_{0i}-F_{ki}N^{k})(F_{0j}-F_{lj}N^{l})-F_{ij}F^{ij}&\!-&\!\nonumber\\\beta_{0},
-\beta_{1}a_{i}B^{i}-\beta_{2}B_{i}B^{i})
\end{eqnarray}
where
\begin{equation}
F_{\mu\nu}=\partial_{\mu}A_{\nu}-\partial_{\nu}A_{\mu},\quad B^{i}=\frac{1}{2}\frac{\Gamma^{ijk}}{\sqrt{g}}F_{jk},
\end{equation}
with $\Gamma^{ijk}$ the Levi-Civita symbol.\\
If we consider the analysis for the electromagnetic field with the only non-vanishing component $A_{t}(r)$ and $\beta_{\mu}=0 (\mu=0, 1, 2)$, the Maxwell equations can be translated as,
\begin{equation}
\partial_{r}(\sqrt{g}NF^{rt})=0,
\end{equation}
which gives us the solution
\begin{equation}
F^{rt}=\frac{Q_{e}}{\sqrt{g}N}.
\end{equation}
We note the integration constant $Q_{e}$ can be interpreted as the charge of the Lifshitz black
hole. According to Ref. [50], we can set the metric components as
\begin{equation}
N=e^{2f(r)},\quad g_{rr}=\frac{1}{e^{2h(r)}},\quad g_{ii}=e^{2l(r)},\quad N_{a}=0,
\end{equation}
and by substitution of the metric (4) in total form of action ($S=S{g}+S_{m}$), we can obtain
\begin{equation}
f(r)=z\ln r+\frac{1}{2}\ln\xi(r),\quad h=\ln r+\frac{1}{2}\ln\xi(r),\quad l=\ln r.
\end{equation}
Finally, for the form of metric, we have
\begin{equation}
ds^{2}=-r^{2z}\xi(r)dt^{2}+\frac{dr^{2}}{r^{2}\xi(r)}+r^{2}dx_{i}^{2},
\end{equation}
with
\begin{equation}
\xi(r)=1-\frac{M}{r^{d+z}}+\frac{2zQ_{e}^{2}}{d\beta(d-z)}\frac{1}{r^{2d}},
\end{equation}
where $M$ and $Q_{e}$ are integral constants which can be understood as the mass and charge
of the Lifshitz black hole.\\
By combination of Eqs. (6) and (7), we have
\begin{equation}
A_{t}(r)=\frac{Q_{e}}{d-z}(1-(\frac{r_{0}}{r})^{d-z}).
\end{equation}
Now, we focus our attention to the application of $NC$ deformation for this type of $HL$ black hole. Firstly, we can define the general form of Lagrangian for the charged test
particle in different black holes
\begin{equation}
\mathcal{L}=\frac{1}{2}(g_{\mu\nu}\frac{dx^{\mu}}{d\tau}\frac{dx^{\nu}}{d\tau})+eA_{\mu}\frac{dx^{\mu}}{d\tau}.
\end{equation}
We restrict the analysis to the equatorial plane with $\theta=\frac{\pi}{2}$, $d\theta=0$ and $\sin\theta=1$. By considering this point and substituting the metric components (11) in the above equation, we can obtain the Lagrangian for $HL$ black holes as
\begin{equation}
\mathcal{L}=\frac{1}{2}(-r^{2z}\xi(r)\dot{t}^{2}+\frac{\dot{r}^{2}}{r^{2}\xi(r)}+r^{2}\dot{x_{i}}^{2})+eA_{t}(r)\dot{t}.
\end{equation}
It is obvious that by the canonical relations, we can attain the corresponding Hamiltonian. In order to apply noncommutative
geometry to the mentioned Hamiltonian, we need to write the Hamiltonian in terms
of new variable. In that case, the new Hamiltonian corresponding to the Lagrangian of $HL$ black holes will be a form of simple harmonic oscillator. So, this form of Hamiltonian leads us to apply non-commutative geometry. First we try to choose following variables
\begin{equation}
\begin{array}{ccc}
x_{1}=r^{z}\sqrt{\xi(r)}\cosh t, & x_{2}=r\sinh x_{i}, & x_{3}=\frac{1}{r\sqrt{\xi(r)}}\sinh r, \\
y_{1}=r^{z}\sqrt{\xi(r)}\sinh t, & y_{2}=r\cosh x_{i}, & y_{3}=\frac{1}{r\sqrt{\xi(r)}}\cosh r,
\end{array}
\end{equation}
\begin{equation}
\begin{array}{cc}
x_{4}+y_{4}=\sqrt{2}r+r^{z}\sqrt{\xi(r)}, & x_{5}+y_{5}=\frac{1}{r\sqrt{\xi(r)}}+r,\\
x_{4}-y_{4}=\sqrt{2}r-r^{z}\sqrt{\xi(r)}, & x_{5}-y_{5}=\frac{1}{r\sqrt{\xi(r)}}-r,
\end{array}
\end{equation}
and
\begin{equation}
\begin{array}{cc}
x_{6}=\cosh\sqrt{eA_{t}(r)\dot{t}}, & x_{7}+y_{7}=\sqrt{eA_{t}(r)\dot{t}}+1,\\
y_{6}=\sinh\sqrt{eA_{t}(r)\dot{t}}, & x_{7}-y_{7}=\sqrt{eA_{t}(r)\dot{t}}-1.
\end{array}
\end{equation}
The Hamiltonian
\begin{equation}
H=\frac{1}{2}\sum\limits_{i=1}^n ((P_{x_{i}}^{2}-P_{y_{i}}^{2})+\omega_{i}^{2}(x_{i}^{2}-y_{i}^{2})),
\end{equation}
where $n=7$
\begin{equation}
\begin{array}{c}
P_{x_{i}}=\frac{\partial\mathcal{L}}{\partial\dot{x_{i}}}=\dot{x_{i}},\\
P_{y_{i}}=\frac{\partial\mathcal{L}}{\partial\dot{y_{i}}}=-\dot{y_{i}}
\end{array}
\end{equation}
and
\begin{eqnarray}
\omega_{i}^{2}=-1,\quad i=1, 2, ..., 5 & \omega_{i}^{2}=0,\quad i=6, 7.
\end{eqnarray}
We can see that the Eq. (19) has the same form with the Hamiltonian of harmonic oscillator as a well-defined system in gravitational
theories. Before application of $NC$ geometry in our set up, it is nice to review some important points about this geometry.\\
The Poisson brackets in commutative case can be found as,
\begin{equation}
\{x_{i},x_{j}\}=0,\quad\{P_{x_{i}},P_{x_{j}}\}=0,\quad\{x_{i},P_{x_{j}}\}=\delta_{ij},
\end{equation}
where $x_{i}(i=1, 2)$ and $P_{x_{i}}(i=1, 2)$. In order to compare the commutative and $NC$ geometries, we explain some points of view about non-commutativity. Quantum effects can be dissolved by the Moyal brackets
$\{f,g\}_{\alpha}=f\star_{\alpha}g-g\star_{\alpha}f$ which is based
on the Moyal product as,
\begin{equation}
(f\star_{\alpha}g)(x)=exp[\frac{1}{2}\alpha^{ab}\partial_{a}^{(1)}\partial_{b}^{(2)}]f(x_{1})g(x_{2})|_{x_{1}=x_{2}=x}.
\end{equation}
The algebra of variables can be driven as,
\begin{equation}
\{x_{i},x_{j}\}_{\alpha}=\theta_{ij},\quad\{x_{i},P_{j}\}_{\alpha}=\delta_{ij}+\sigma_{ij},\quad\{P_{i},P_{j}\}=\beta_{ij}.
\end{equation}
Transformations on the classical phase space variables are expressed as,
\begin{equation}
\hat{x_{i}}=x_{i}+\frac{\theta}{2}P_{y_{i}},\quad \hat{y_{i}}=y_{i}-\frac{\theta}{2}P_{x_{i}},\quad \hat{P_{x_{i}}}=P_{x_{i}}-\frac{\beta}{2}y_{i},
\quad \hat{P_{y_{i}}}=P_{y_{i}}+\frac{\beta}{2}x_{i},\\
\end{equation}
The algebra for new variables are given by,
\begin{equation}
\{\hat{y},\hat{x}\}=\theta,\quad\{\hat{x},\hat{P_{x}}\}=\{\hat{y},\hat{P_{y}}\}=1+\sigma,\quad\{\hat{P_{y}},\hat{P_{x}}\}=\beta,
\end{equation}
where $\sigma=\frac{\beta\theta}{2}$. By rewriting the Eq. (19) with new variables in Eq. (25), we can access
to the deformed form of charged test particle in $HL$ black holes as,
\begin{equation}
\hat{H}=\frac{1}{2}\sum_{i=1}^{n}((P_{x_{i}}^{2}-P_{y_{i}}^{2})-\gamma_{i}^{2}(y_{i}P_{x_{i}}+x_{i}P_{y_{i}})+
{\tilde{\omega_{i}}^{2}}(x_{i}^{2}-y_{i}^{2})),
\end{equation}
where
\begin{equation}
{\tilde{\omega_{i}}}^{2}=\frac{{\omega_{i}}^{2}-\frac{\beta^{2}}{4}}{1-{\omega_{i}}^{2}
\frac{\theta^{2}}{4}},\quad
{\gamma_{i}}^{2}=\frac{\beta-{\omega_{i}}^{2}\theta}{1-{\omega_{i}}^{2}\frac{\theta^{2}}{4}}.
\end{equation}
Also, the form of Lagrangian for the above Hamiltonian can be expressed as, \begin{equation}
\hat{\mathcal{L}}=\frac{1}{2}\sum_{i=1}^{n}((P_{x_{i}}^{2}-P_{y_{i}}^{2})-{\hat{\gamma_{i}}^{2}}(y_{i}P_{x_{i}}+x_{i}P_{y_{i}})-
{\hat{\omega_{i}}^{2}}(x_{i}^{2}-y_{i}^{2})),
\end{equation}
where
\begin{equation}
{\hat{\omega_{i}}}^{2}=\frac{{\omega_{i}}^{2}+\frac{\beta^{2}}{4}}{1+{\omega_{i}}^{2}
\frac{\theta^{2}}{4}},\quad
{\hat{\gamma_{i}}}^{2}=\frac{\beta+{\omega_{i}}^{2}\theta}{1+{\omega_{i}}^{2}\frac{\theta^{2}}{4}}.
\end{equation}
We use the Eq. (20), then the deformed Lagrangian can be written as following,
\begin{eqnarray}
\hat{\mathcal{L}}=\frac{1}{2}[-r^{2z}\xi(r)\dot{t}^{2}+\frac{\dot{r}^{2}}{r^{2}\xi(r)}+r^{2}\dot{x_{i}}^{2}+
(\frac{(\beta-\theta)}{1-\frac{\theta^{2}}{4}}r^{2z}\xi(r)+\frac{\beta^{2}}{4}eA_{t}(r))\dot{t}-\frac{(\beta-\theta)}{1-\frac{\theta^{2}}{4}}(r^{2}\dot{x_{i}}&\!+&\!\nonumber\\
(\frac{1}{r^{2}\xi(r)}+\sqrt{2}(1-z)r^{z}\sqrt{\xi(r)}-\frac{\sqrt{2}r^{z+1}\acute{\xi}(r)}{2\sqrt{\xi(r)}}
-\frac{2}{r\sqrt{\xi(r)}}-\frac{\acute{\xi}(r)}{2\xi(r)\sqrt{\xi(r)}})\dot{r})],
\end{eqnarray}
where $\acute{\xi}(r)=\frac{d\xi(r)}{dr}$. In order to find $V_{eff}$ for this $HL$ black hole, we can present the following strategy. One can define the conjugate momentum $P_{\mu}$ to the coordinate $x^{\mu}$ as
\begin{equation}
P_{\mu}\equiv\frac{\partial \mathcal{L}}{\partial\dot{x}^{\mu}}=g_{\mu\nu}\dot{x}^{\nu},
\end{equation}
where
\begin{equation}
\dot{x}^{\mu}\equiv\frac{dx^{\mu}}{d\tau}=u^{\mu}.
\end{equation}
In terms of the conjugate momenta, the Euler-Lagrange equations can be written as
\begin{equation}
 \frac{d}{d\tau}P_{\mu}=\frac{\partial\mathcal{L}}{\partial x^{\mu}}.
\end{equation}
The mentioned black hole is stationary and axisymmetric, so there are two killing vectors $k^{\mu}=(1,0,0,0)$ and $m^{\mu}=(0,0,0,1)$
which are time-like and space-like, respectively. In order to obtain the conserved constants, one can find the geodesics equations by the above Killing vectors. Then, for constants, we can find $E=-\partial_t. U$ and $L=\partial_\phi. U$
where $U$, $E$ and $L$ correspond to the four-velocity and its energy and angular
momentum in our system. In this work, we restrict ourselves to the equatorial plane $(\theta =\frac{\pi}{2})$ implying
$U^\theta=0 $. The radial geodesic is obtained by solving the equation
$U.U=-1.$ Therefore, the canonical momenta $P_t$ and $P_{\phi}$ will be conserved which $E$ and $L$ are interpreted as energy and angular moment per unit mass, respectively as,
\begin{equation}
E\equiv -k_{\mu}u^{\mu}=-g_{t\mu}u^{\mu}=-P_{t},
\end{equation}
\begin{equation}
L\equiv m_{\mu}u^{\mu}=-g_{\varphi\mu}u^{\mu}=P_{\varphi},
\end{equation}
where $E$ is the energy at infinity and $L$ is the angular momentum. In order to obtain $\dot{t}$ and $\dot{\varphi}$, we have to consider Eqs. (15) and (34), so we obtain the following equation,
\begin{equation}
\dot{t}=-(\frac{K_{1}}{r^{2z}\xi(r)}+\frac{K_{2}}{r^{d+z}\xi(r)})u^{t},
\end{equation}
where
\begin{equation}
K_{1}=E-\frac{eQ_{e}}{d-z}, \quad K_{2}=\frac{eQ_{e}}{d-z}r_{0}^{d-z}
\end{equation}
\begin{equation}
\dot{\varphi}=\frac{L}{r^{2}}=u^{\varphi}.
\end{equation}
Furthermore, the subject of effective potential and center mass energy play an important role in collision of two particles on the black hole background. In order to have information about such quantities, we employ the following equation,
\begin{equation}
g_{\mu\nu}u^{\mu}u^{\nu}=\kappa,
\end{equation}
where $\kappa=-1$ for time-like geodesics. In that case, for obtaining effective potential we use Eqs. (37) to (40) and $\dot{r}^{2}+V_{eff}=0$. So $V_{eff}$ can be written by following equation,
\begin{equation}
V_{eff}=r^{2}\xi(r)(\frac{L^{2}}{r^{2}}+1)-(\frac{K_{1}}{r^{z-1}}+\frac{K_{2}}{r^{d-1}})^{2}.
\end{equation}
Now, we can concentrate our attention on Finsler geometry issues and also we take advantage of Lagrangian in charged test particle in $HL$ black holes. At the first, we are going to review some properties in Finsler geometry. As we said a Finsler space is an $N$-dimensional manifold $M$
equipped with a Finsler metric which is equipped with some special conditions.\\
Now, we can introduce the general form of metric in Finsler geometry which is given by following,
\begin{equation}
 F(x,y)=\alpha a(x,y)+\eta b(x,y)+\gamma \frac{a^{2}(x,y)}{b(x,y)}+\Xi(x,y),
\end{equation}
 where $a(x,y)=\sqrt{a_{ij}(x)y^{i}y^{j}}$ is a Riemannian
metric, $a^{ij}$ is the
inverse matrix of $a_{ij}$  and $b(x,y)$ is a differential one-form on $M$ with
$\|b(x,y)\|_{a}:=\sqrt{a^{ij}b_{i}b_{j}}<1$. Also one can say that
$\Xi(x,y)$ is a Finsler fundamental function on $TM$  and $\alpha, \eta, \gamma\in\mathcal{F}(M)$. The first two terms of $F$  in above metric determine a Randers metric. The Finsler metric (Lagrangian) corresponding to charged test particle in $HL$ black holes will be as,
\begin{eqnarray}
F(x,y)=\alpha\sqrt{-r^{2z}\xi(r)\dot{t}^{2}+\frac{\dot{r}^{2}}{r^{2}\xi(r)}+r^{2}\dot{x_{i}}^{2}}+
(\alpha_{1}r^{2z}\xi(r)+\alpha_{2}eA_{t}(r))\dot{t}+\alpha_{3}r^{2}\dot{x_{i}}&\!+&\!\nonumber\\
\alpha_{4}(\frac{1}{r^{2}\xi(r)}+\sqrt{2}(1-z)r^{z}\sqrt{\xi(r)}-\frac{\sqrt{2}r^{z+1}\acute{\xi}(r)}{2\sqrt{\xi(r)}}
-\frac{2}{r\sqrt{\xi(r)}}-\frac{\acute{\xi}(r)}{2\xi(r)\sqrt{\xi(r)}})\dot{r}.
\end{eqnarray}
The above equation can be obtained Eq. (42) if $\Xi=\gamma=0$. The most important result here is the comparison of obtained results of Lagrangian between Finsler and $NC$ geometry. In order to compare such results in Eq. (31), we have to apply following conditions to Eq. (43)
\begin{equation}
 \dot{r}^{2} \ll V_{eff},\qquad \alpha=\sqrt{\xi(r)L^{2}-(\frac{K_{1}}{r^{z-1}}+\frac{K_{2}}{r^{d-1}})^{2}}, \quad \eta=1.
\end{equation}
Also, we note that the system with mentioned conditions has a stability as ${\omega_{\varphi}}^{2} \gg 1$.
According to the mentioned conditions $F(x,y)$ can be written by,
\begin{eqnarray}
F(x,y)\simeq-r^{2z}\xi(r)\dot{t}^{2}+\frac{\dot{r}^{2}}{2r^{2}\xi(r)}+r^{2}\dot{x_{i}}^{2}+
(\alpha_{1}r^{2z}\xi(r)+\alpha_{2}eA_{t}(r))\dot{t}+\alpha_{3}r^{2}\dot{x_{i}}&\!+&\!\nonumber\\
\alpha_{4}(\frac{1}{r^{2}\xi(r)}+\sqrt{2}(1-z)r^{z}\sqrt{\xi(r)}-\frac{\sqrt{2}r^{z+1}\acute{\xi}(r)}{2\sqrt{\xi(r)}}.
-\frac{2}{r\sqrt{\xi(r)}}-\frac{\acute{\xi}(r)}{2\xi(r)\sqrt{\xi(r)}})\dot{r}
\end{eqnarray}
As a consequence of the above comparison, $\alpha_{1},$ $\alpha_{2}$, $\alpha_{3}$ and $\alpha_{4}$ are depended to the NC parameters $\beta$ and $\theta$.
\subsection{The non-charged Horava-Lifshitz black hole}
In the previous section, by putting $Q_{e}=0$ in the Lagrangian (14), we can attain the Lagrangian of the non-charge $HL$ black holes. Our analysis can be repeated for this black hole as follow. The corresponding Hamiltonian of the above Lagrangian can be obtained by definition of the new variables. For this kind of $HL$ black hole, the variables are the similar to the previous black hole without $x_{6}$, $x_{7}$, $y_{6}$ and $y_{7}$. Also, the Hamiltonian has the same form with the Eq. (19) with $n=5$. For application of $NC$ deformation, we can engage the variable (16) and (17). Finally, we can find that the deformed Hamiltonian and then the deformed Lagrangian are similar to the charged $HL$ black hole (27) and (29) as with $n=5$. Also, in order to attain the effective potential in this kind of $HL$ black hole, we can neglect the role of charge in the Eq. (41). On the other hand, in Finsler regime, the results are again similar to the previous black hole if we remove charge from our calculations. Therefore, for this black hole the Eqs. (43) to (45) are valid for the case of $A_{\mu}=0$.
\subsection{The non-rotated Horava-Lifshitz black hole}
According to the physics of black holes, we cannot neglect the role of rotation in our investigation. But, in this section we apply the analysis for non-rotated $HL$ black holes without charge property. The general form of action in this case, can be expressed as [51],
\begin{equation}
S=\int dtdx^{3}\sqrt{g}\bar{N}[\bar{\mathcal{L}}_{0}+\mathcal{L}_{0}+\mathcal{L}_{1}],
\end{equation}
where
\begin{eqnarray}
˜\bar{\mathcal{L}}_{0}=\frac{2}{k^{2}}(K_{ij}K^{ij}-\lambda K^{2}),\nonumber\hspace{5.95cm}\\
\mathcal{L}_{0}=-\frac{k}{2\omega^{4}}C_{ij}C^{ij}\frac{k^{2}\mu}{2\omega^{2}}\epsilon^{ijk}R^{(3)}_{il} \nabla_{j}R_{k}^{(3)l}-\frac{k^{2}\mu^{2}}{8}R_{ij}^{(3)}R^{(3)ij},\nonumber\hspace{0.7cm}\\
\mathcal{L}_{1}=\frac{k^{2}\mu^{2}}{8(1-3\lambda)}(\frac{1-4\lambda}{4}(R^{(3)})^{2}+\Lambda_{W}R^{(3)}-3\Lambda^{2}_{W})
+\mu^{4}R^{(3)},
\end{eqnarray}
where k, $\lambda$ and $\mu$ are constant parameters. Also, we can find the formula $C^{ij}=\epsilon^{ijk}\nabla_{k}(R_{l}^{j}-\frac{1}{4}R\delta_{l}^{j})$ and $K_{ij}\frac{1}{2\bar{N}}=\dot{g}_{ij}-\nabla_{i}N_{j}-\nabla_{j}N_{i}$ where $\bar{N}$ and $N_{i}$ are the lapse and shift functions respectively.\\
In order to obtain the black hole solution, we can define a general form of metric as follow
\begin{equation}
ds^{2}=f(r)dt^{2}-\frac{dr^{2}}{f(r)}-r^{2}(d\theta^{2}+\sin^{2}\theta d\phi^{2}),
\end{equation}
where for this kind of black hole, we have
\begin{equation}
f(r)= 1+(\omega-\Lambda_{W})r^{2}-r[\omega(\omega-2\Lambda_{W})r^{3}+\beta])^{\frac{1}{2}}.
\end{equation}
Here, $\beta$ is integration constant, $\Lambda_{W}$ and $\omega$ are real constant parameters. In the analysis, we consider the case of Kehagias-Sfetsos $(KS)$ black hole solution with $\beta=4\omega M$ and $\Lambda_{W}=0$, so the form of $f_(r)$ can be written as
\begin{equation}
f_{KS}(r)=1+\omega r^{2}-\omega r^{2}\sqrt{1+\frac{4M}{\omega r^{3}}}.
\end{equation}
To engage $NC$ geometry for this black hole firstly, we consider the Eq. (14) for the case of $A_{\mu}=0$ and then use the metric (48) in equatorial plane ($\theta=\frac{\pi}{2}$). So, the Lagrangian can be found as
\begin{equation}
\mathcal{L}=-\frac{1}{2}(f(r)\dot{t}^{2}-\frac{\dot{r}^{2}}{f(r)}-r^{2}\dot{\phi}^{2}).
\end{equation}
By defining the new following variables,
\begin{equation}
\begin{array}{ccc}
x_{1}=\sqrt{f(r)}\cosh t, & x_{2}=r\sinh\phi, & x_{3}=\frac{1}{\sqrt{f(r)}}\sinh r,\\
y_{1}=\sqrt{f(r)}\sinh t, & y_{2}=r\cosh\phi, & y_{3}=\frac{1}{\sqrt{f(r)}}\cosh r,\\
\end{array}
\end{equation}
\begin{equation}
\begin{array}{cc}
x_{4}+y_{4}=\sqrt{2}r+\sqrt{f(r)}, & x_{5}+y_{5}=\frac{1}{\sqrt{f(r)}}+r,\\
x_{4}-y_{4}=\sqrt{2}r-\sqrt{f(r)}, & x_{5}-y_{5}=\frac{1}{\sqrt{f(r)}}-r.
\end{array}
\end{equation}
and by regarding to the our strategy in the previous section for $n=5$, the shape of $NC$ deformed Lagrangian can be driven as,
\begin{eqnarray}
\hat{\mathcal{L}}=\frac{1}{2}[-(f(r)\dot{t}^{2}+\frac{1}{f(r)}\dot{r}^{2}+r^{2}\dot{\phi}^{2}-
\frac{\beta-\theta}{1-\frac{\theta^{2}}{4}}(-f(r)\dot{t}+r^{2}\dot{\phi}+(\frac{1}{f(r)}&\!+&\!\nonumber\\
\sqrt{2f(r)}-\frac{1}{\sqrt{f(r)}}-\frac{\acute{f}(r)(1+\sqrt{2}f(r))}{2f(r)\sqrt{f(r)}}r)\dot{r})].
\end{eqnarray}
The effective potential for this black hole can be found by using the Eqs. (34) and (51)
\begin{equation}
V_{eff}=f(r)(\frac{L^{2}}{r^{2}}+1)-E^{2}.
\end{equation}
Also, the form of Finsler metric (42) in this kind of $HL$ black hole can be found as,
\begin{eqnarray}
F(x,y)=\alpha\sqrt{-f(r)\dot{t}^{2}+\frac{1}{f(r)}\dot{r}^{2}+r^{2}\dot{\phi}^{2}}+(-\alpha_{1}f(r)\dot{t}+\alpha_{2}r^{2}\dot{\phi}+\alpha_{3}
(\frac{1}{f(r)}&\!+&\!\nonumber\\
\sqrt{2f(r)}-\frac{1}{\sqrt{f(r)}}-\frac{\acute{f}(r)(1+\sqrt{2}f(r))}{2f(r)\sqrt{f(r)}}r)\dot{r}).
\end{eqnarray}
According to the following conditions
\begin{equation}
\dot{r}^{2} \ll V_{eff},\,\,\alpha=\sqrt{\frac{E^{2}}{f(r)}-\frac{L^{2}}{r^{2}}},\,\,\eta=1.
\end{equation}
and the stability condition ${\omega_{\varphi}}^{2} \gg 1$, the Finsler metric is obtained as,
\begin{eqnarray}
F(x,y)\simeq-f(r)\dot{t}^{2}+\frac{1}{2}\frac{1}{f(r)}\dot{r}^{2}+r^{2}\dot{\phi}^{2}+
(-\alpha_{1}f(r)\dot{t}+\alpha_{2}r^{2}\dot{\phi}+\alpha_{3}(\frac{1}{f(r)}&\!+&\!\nonumber\\
\sqrt{2f(r)}-\frac{1}{\sqrt{f(r)}}-\frac{\acute{f}(r)(1+\sqrt{2}f(r))}{2f(r)\sqrt{f(r)}}r)\dot{r}).
\end{eqnarray}
By comparing the Eqs. (54) and (58), we can see similarity between Finsler and $NC$ metrics.
\subsection{The rotated Horava-Lifshitz black hole}
In the present work, another interesting black hole is rotated $HL$ black hole. The general form of metric for this black hole can be expressed as
\begin{equation}
ds^{2}=f(r)dt^{2}-\frac{dr^{2}}{f(r)}-r^{2}(d\theta^{2}-r^{2}\sin^{2}\theta(d\phi-a Ndt)^{2},
\end{equation}
where
\begin{equation}
N=\frac{2M}{r^{3}}.
\end{equation}
Also, if we restrict our analysis for slowly $KS$ black hole solution, the metric can be reduced to
\begin{equation}
ds^{2}_{slow KS}=f_{KS}(r)dt^{2}-\frac{dr^{2}}{f_{KS}(r)}-r^{2}d\phi^{2}+\frac{4J}{r}dtd\phi.
\end{equation}
In order to apply $NC$ geometry for this black hole, we need to find the form of corresponding Lagrangian. So, by using Eq. (61) in Eq. (14) with $A_{\mu}=0$ for equatorial plane, the Lagrangian can be driven as
\begin{equation}
\mathcal{L}=\frac{1}{2}(f_{KS}(r)\dot{t}^{2}-\frac{\dot{r}^{2}}{f_{KS}(r)}-r^{2}\dot{\phi}^{2}+
\frac{4J}{r}\dot{t}\dot{\phi}).
\end{equation}
By definition of new variables,
\begin{equation}
\begin{array}{ccc}
x_{1}=\sqrt{f_{KS}(r)}\sinh t, & x_{2}=r\cosh\phi, & x_{3}=\frac{1}{\sqrt{f_{KS}(r)}}\cosh r,\\
y_{1}=\sqrt{f_{KS}(r)}\cosh t, & y_{2}=r\sinh\phi, & y_{3}=\frac{1}{\sqrt{f_{KS}(r)}}\sinh r,\\
\end{array}
\end{equation}
and
\begin{equation}
\begin{array}{cc}
x_{4}+y_{4}=\sqrt{f_{KS}(r)}+\sqrt{2}r, & x_{5}+y_{5}=r+\frac{1}{\sqrt{f_{KS}(r)}},\\
x_{4}-y_{4}=\sqrt{f_{KS}(r)}-\sqrt{2}r, & x_{5}-y_{5}=r-\frac{1}{\sqrt{f_{KS}(r)}}.
\end{array}
\end{equation}
\begin{equation}
\begin{array}{ccc}
x_{6}=\frac{J}{r}t+\frac{1}{2}\phi, & x_{7}=\frac{1}{2}t+\frac{J}{r}\phi, & x_{8}=t\phi-\frac{J}{2r},\\
y_{6}=\frac{J}{r}t-\frac{1}{2}\phi, & y_{7}=\frac{1}{2}t-\frac{J}{r}\phi, & y_{8}=t\phi+\frac{J}{2r},\\
\end{array}
\end{equation}
and by application of the Eqs. (19) and (20) for $n = 8$ and $w_{i}^{2}=-1,\,\, i=1,2,...,7$ and $w_{8}^{2}=-2$, the shape of deformed Lagrangian takes the following form
\begin{eqnarray}
\hat{\mathcal{L}}=\frac{1}{2}[(-f_{KS}(r)\dot{t}^{2}+\frac{1}{f_{KS}(r)}\dot{r}^{2}+r^{2}\dot{\phi}^{2}+\frac{4J}{r}\dot{t}\dot{\phi})
+((\theta-\beta)f_{KS}(r)+\beta\frac{J\phi}{r})\dot{t}+((\beta-\theta)r^{2}&\!+&\!\nonumber\\(4\theta-3\beta)\frac{Jt}{r})\dot{\phi}+
((\theta-\beta)(-\sqrt{2f_{KS}(r)}-\frac{1}{f_{KS}(r)}+\frac{\sqrt{2}}{2}\frac{\acute{f}_{KS}(r)r}{\sqrt{f_{KS}(r)}}
+\frac{2f_{KS}(r)-\acute{f}_{KS}(r)r}{2f_{KS}(r)\sqrt{f_{KS}(r)}})&\!+&\!\nonumber\\(2\theta-\beta)\frac{Jt\phi}{r^{2}})\dot{r}].
\end{eqnarray}
In order to obtain the form of effective potential for this black hole, we use the Eqs. (62) and (34). Then, $V_{eff}$ can be found
\begin{equation}
V_{eff}=A[1-\frac{C}{\triangle}(E-V_{+})(E-V_{-})],
\end{equation}
where $A=-f_{KS}(r)$, $B=\frac{4J}{r}$, $C=-r^{2}$ and $\triangle=AC+B^{2}$. Also, $V_{\pm}$ takes the following form,
\begin{equation}
V_{\pm}=\frac{-4JL\mp|L|\sqrt{f_{KS}(r)r^{4}+16J^{2}}}{r^{3}}.
\end{equation}
For this black hole, the Finsler metric (42) can be expressed as,
\begin{eqnarray}
F(x,y)=\alpha\sqrt{(-f_{KS}(r)\dot{t}^{2}+\frac{1}{f_{KS}(r)}\dot{r}^{2}+r^{2}\dot{\phi}^{2}+\frac{4J}{r}\dot{t}\dot{\phi})}
+(\alpha_{1}f_{KS}(r)+\alpha_{2}\frac{J\phi}{r})\dot{t}+(\alpha_{3}r^{2}&\!+&\!\nonumber\\
\alpha_{4}\frac{Jt}{r})\dot{\phi}+(\alpha_{5}(-\sqrt{2f_{KS}(r)}-\frac{1}{f_{KS}(r)}+\frac{\sqrt{2}}{2}\frac{\acute{f}_{KS}(r)r}{\sqrt{f_{KS}(r)}}
+\frac{2f_{KS}(r)-\acute{f}_{KS}(r)r}{2f_{KS}(r)\sqrt{f_{KS}(r)}})&\!+&\!\nonumber\\
\alpha_{6}\frac{Jt\phi}{r^{2}})\dot{r},
\end{eqnarray}
and by considering such conditions,
\begin{equation}
\dot{r}^{2} \ll A-V_{eff},\,\,\,\alpha=\sqrt{\frac{C}{\Delta}(E-V_{+})(E-V_{+})},\,\,\,\eta=1.
\end{equation}
This equation lead us to have $0<r<r_H$ , this condition completely guarantee to have following expression for the Finsler geometry,
\begin{eqnarray}
F(x,y)\simeq(-f_{KS}(r)\dot{t}^{2}+\frac{1}{2}\frac{1}{f_{KS}(r)}\dot{r}^{2}+r^{2}\dot{\phi}^{2}+\frac{4J}{r}\dot{t}\dot{\phi})
+(\alpha_{1}f_{KS}(r)+\alpha_{2}\frac{J\phi}{r})\dot{t}+(\alpha_{3}r^{2}&\!+&\!\nonumber\\
\alpha_{4}\frac{Jt}{r})\dot{\phi}+
(\alpha_{5}(-\sqrt{2f_{KS}(r)}-\frac{1}{f_{KS}(r)}+\frac{\sqrt{2}}{2}\frac{\acute{f}_{KS}(r)r}{\sqrt{f_{KS}(r)}}
+\frac{2f_{KS}(r)-\acute{f}_{KS}(r)r}{2f_{KS}(r)\sqrt{f_{KS}(r)}})&\!+&\!\nonumber\\\alpha_{6}\frac{Jt\phi}{r^{2}})\dot{r}.
\end{eqnarray}
By applying the Eq. (70) and $0<r<r_H$ in Eq. (69), one can see that the Eqs. (66) and (71) will be the same. Such similarity navigate us to connect parameters from two geometries as $\alpha_{1},$ to $\alpha_{6}$ are depended to $NC$ parameters.
\section{Conclusion}
In the present paper, we focused on a class of $HL$ black holes in the context of $HL$ gravity as a candidate to explain $QG$. Then, we carried out the analysis in two geometrical approaches ($NC$ and Finsler geometry) for four kinds of $HL$ black hole as charged, non-charged, non-rotated and rotated $HL$ black holes. In order to obtain our main purpose, we compared the form of Lagrangian in two geometries together for the mentioned $HL$ black holes. As a consequence of this comparison, we found that there is a similarity between  form of Lagrangian in two geometries under some conditions. In fact, it seems that these geometries are equivalent in specific situations. The similarity between two geometries about the Lagrangian of four kind of $HL$ black hole and Ref. [31] lead us to think about universal rule for all of black holes. We note that this equivalence is obtained only for $HL$ black holes and it will be interesting to generalize this approach to other cosmological scenarios and different metric background.


\begin{thebibliography}{}
\bibitem{1} C.J. Isham, [arXiv:gr-qc/9310031].
\bibitem{2} S. Carlip, \textit{Rept. Prog. Phys} \textbf{64} 885 (2001).
\bibitem{3} J. M. Maldacena, \textit{Adv. Theor. Math. Phys.} \textbf{2} 231 (1998).
\bibitem{4} S. S. Gubser, I. R. Klebanov and A. M. Polyakov, \textit{Phys. Lett. B} \textbf{428} 105 (1998).
\bibitem{5} E. Witten, \textit{Adv. Theor. Math. Phys.} \textbf{2} 253 (1998).
\bibitem{6} B. Schulz, [arxiv: gr-qc/1409.7977].
\bibitem{7} P. Horava, \textit{Phys. Rev. Lett.} \textbf{102} 161301 (2009).
\bibitem{8} P. Horava, \textit{Phys. Rev. D} \textbf{79} 084008 (2009).
\bibitem{9} J. Rembielinski, \textit{Phys. Lett. B} \textbf{67} 730 (2014).
\bibitem{10} T. Griffin, P. Horava and CH. M. Melby-Thompsonc, \textit{Phys. Rev. Lett.} \textbf{110} 081602 (2013).
\bibitem{11} C. Hoyos, B. S. Kim and Y. Oz, \textit{JHEP} \textbf{11} 145 (2013).
\bibitem{12} S. Janiszewski, A. Karch, B. Robinson and D. Sommer, \textit{JHEP} \textbf{04} 163 (2014).
\bibitem{13} F. W. Shu, K. Lin, A. Wang and Q. Wu, \textit{JHEP} \textbf{04} 056 (2014).
\bibitem{14} Ch. Eling and Y. Oz, \textit{JHEP} \textbf{11} 067 (2014).
\bibitem{15} A. Borzou, K. Lin and A. Wang, \textit{JCAP} \textbf{02} 025 (2012).
\bibitem{16} E. Kiritsis and G. Kofinas, \textit{JHEP} \textbf{01} 122 (2010).
\bibitem{17} A. Wang, \textit{Int. J. Mod. Phys. D} \textbf{26}	1730014 (2017).
\bibitem{18} Y. S. Myung, \textit{Phys. Lett. B} \textbf{690} 534 (2010).
\bibitem{19} R.B. Mann, \textit{JHEP} \textbf{0906} 075 (2009).
\bibitem{20} K. Balasubramanian and J. McGreevy, \textit{Phys. Rev. D} \textbf{80} 104039 (2009).
\bibitem{21} G. Bertoldi, B. A. Burrington and A. W. Peet, \textit{Phys. Rev. D} \textbf{80} 126003 (2009).
\bibitem{22} G. Bertoldi, B. A. Burrington and A. W. Peet, \textit{Phys. Rev. D} \textbf{80} 126004 (2009).
\bibitem{23} E. Ayon-Beato, A. Garbarz, G. Giribet and M. Hassaine, [arXiv:hep-th/1001.2361].
\bibitem{24} E. Ayon-Beato, A. Garbarz, G. Giribet and M. Hassaine, \textit{Phys. Rev. D} \textbf{80} 104029 (2009).
\bibitem{25} E .A. Bergshoeff, O. Hohm and P. K. Townsend, \textit{Phys. Rev. Lett.} \textbf{102} 201301 (2009).
\bibitem{26} R. G. Cai, L. M. Cao and N. Ohta, \textit{Phys. Lett. B} \textbf{679} 504 (2009).
\bibitem{27} Kh. Jafarzade and J. Sadeghi, [arXiv:hep-th/1604.02973].
\bibitem{28} A. Connes, \textit{Noncommutative Geometry}, Academic Press Inc. San Diego (1994).
\bibitem{29} N. Seiberg and E. Witten, \textit{JHEP} \textbf{9909} 032 (1999).
\bibitem{30} J. Sadeghi, B. Pourhassan, Z. Nekouee and M. Shokri, \textit{Int. J. Mod. Phys. D} \textbf{27} 1850025 (2018).
\bibitem{31} J. Sadeghi, Z. Nekouee and A. Behzadi, [arXiv:hep-th/1710.01568].
\bibitem{32} J. Sadeghi, Z. Nekouee and M. Shokri, [arXiv:hep-th/1711.02534].
\bibitem{33} G. S. Asanov, \textit{Lett. Nuovo Cimento}, \textbf{221} 49 (1979).
\bibitem{34} G. Yu. Bogolovsky, \textit{Lett. Nuovo Cimento} \textbf{99} 40 (1977).
\bibitem{35} T. Ishikawa, \textit{J. Math. Phys.} \textbf{995} 22 (1981).
\bibitem{36} M. Matsumato, \textit{Rep. Math. Phys.} \textbf{103} 8 (1975).
\bibitem{37} Y. Tokano,\textit{ Lett. Nuovo Cimento} \textbf{307} 10 (1974).
\bibitem{38} I. W. Roxburgh,\textit{ Gen. Rel. Grav. }\textbf{419} 24 (1992).
\bibitem{39} I. W. Roxburgh and R. K. Takavol, \textit{Gen. Rel. Grav.} \textbf{307} 10 (1979).
\bibitem{40} R. K. Takavol and N. Van der Bergh, \textit{Gen. Rel. Grav.} \textbf{849} 18 (1986).
\bibitem{41} S. F. Rutz, \textit{Gen. Rel. Grav.} \textbf{1139} 25 (1993).
\bibitem{42} D. Bao, S. S. Chern and Z. Shen, (Springer, New York, 2000).
\bibitem{43} S. S. Chern, W. H. Chen and K. S. Lam, \textit{Lectures on Differential Geometry}, (World Scientific, 2000).
\bibitem{44} L. Kozma and T. Ootsuka,\textit{ Rep. Math. Phys.} \textbf{157} 78 (2016) .
\bibitem{45} Y. Suzuki, \textit{Journal of the College of Arts and Sciences}, Chiba University \textbf{12} 2 (1956) .
\bibitem{46} C. Lanczos, \textit{The Variational Principles of Mechanics}, (Dover Books on Physics, 1986).
\bibitem{47} L. Kozma and L. Tamassy, \textit{Rep. Math. Phys.} \textbf{233} 51 (2003).
\bibitem{48} T. Griffin, P. Horava and C. M. Melby-Thompson, \textit{Phys. Rev. Lett} \textbf{110} 081602 (2013).
\bibitem{49} I. Kimpton and A. Padilla, \textit{JHEP} \textbf{04} 133 (2013).
\bibitem{50} M. Alishahiha and H. Yavartanoo, \textit{Class. Quant. Grav.} \textbf{31} 095008 (2014).
\bibitem{51} J. Sadeghi and B. Pourhassan, \textit{Eur. Phys. J. C} \textbf{1984} 72 (2012).
\end{thebibliography}
\end{document}